\documentclass{mem}
\usepackage{natbib}\usepackage{txfonts}\usepackage{balance}
\usepackage{graphicx}
\usepackage[a4paper,breaklinks,dvipdfm]{hyperref}
\idline{75}{282}

\def\saxj{SAX J1808.4-3658}
\def\igra{IGR J00291+5934}
\def\xtea{XTE J1807--294}

\def\ltsima{$\; \buildrel < \over \sim \;$}
\def\simlt{\lower.5ex\hbox{\ltsima}}
\def\gtsima{$\; \buildrel > \over \sim \;$}
\def\simgt{\lower.5ex\hbox{\gtsima}}
\begin{document}
\def\teff{$T\rm_{eff }$}
\def\kms{$\mathrm {km s}^{-1}$}
\title{On Low Mass X-ray Binaries and Millisecond Pulsar}
\author{Luciano Burderi\inst{1,2},Tiziana Di Salvo\inst{3,2}}
\offprints{L. Burderi}
\institute{
Dipartimento di Fisica --
Universit\'a di Cagliari, 
Complesso Universitario di Monserrato, 
S.P. Monserrato-Sestu Km 0,700,
I--09042 Monserrato (Cagliari), Italy
\and
Istituto Nazionale di Astrofisica,
viale del Parco Mellini 84, 
I--00136 Roma, Italy
\and
Dipartimento di Fisca e Chimica --
Universit\'a di Palermo,
via Archirafi 36,
I--90123 Plermo, Italy
\email{burderi@dsf.unica.it}
}
\authorrunning{Burderi}
\titlerunning{On Low Mass X-ray Binaries and Millisecond Pulsar}
\abstract{
The detection, in 1998, of the first Accreting Millisecond Pulsar, 
started an exiciting season of continuing discoveries in the 
fashinating field of compact binary systems harbouring a neutron star.
Indeed, in these last three lustres, thanks to the extraordinary 
performances of astronomical detectors, on ground as well as on board 
of satellites, mainly in the radio, optical, x--ray, and gamma--ray bands,
astrophysicists had the opportunity to thoroughly investigate
the so--called Recycling Scenario: the evolutionary path leading to the 
formation of a Millisecond Radio Pulsar. The most intriguing phase is 
certainly the spin--up stage during which, because of the accretion of matter
and angular momentum, the neutron star accumulates an extraordinary amount
of mechanical rotational energy, up to 1\% of its whole rest--mass energy. 
These millisecond spinning neutron stars are truly extreme physical objects:
General and Special Relativity are fully in action, since their surfaces,
attaining speeds close to one fifth of the speed of light, are located
just beyond their Schwartzscild Radius, and electrodynamical forces,
caused by the presence of huge surface magnetic fields of several 
hundred million Gauss, display their spectacular properties accelerating
electrons up to such energies to promote pair creation in a cascade 
process responsible for the emission in Radio and Gamma--ray.
The rotational energy is swiftly converted and released into 
electromagnetic power which, in some cases, causes the neutron star 
to outshine with a luminosity of one hundred suns.
Along these fifteen years, a fruitful collaboration was established,
at the Rome Astronomical Observatory, between my group and Franca 
D'Antona: her profound knowledge of the complex phases of stellar
evolution, in particular of low--mass stars in close binary systems,
was the key ingredient which boosted our theoretical and experimental
studies of different evolutionary stages of these intriguing and fashinating 
systems. In this paper I will review some of the most 
recent discoveries on (accreting) millisecond pulsars, highlighting 
the role played by our proficuous collaboration.
}
\maketitle{}
\section{Introduction}
The Recycling Scenario explains Neutron Stars (NS hereafter)
spinning at millisecond periods as the result of accretion of
matter and angular momentum from Roche Lobe overflow of a
($\le 1 \ {\rm M_{\odot}}$) companion
(see {\it e.g.} Bhattacharya \& van den Heuvel, 1991, for a review).
According to this model, a subclass of NS Low Mass X-ray Binaries
(LMXBs, hereafter), are the progenitors of the
Millisecond Radio Pulsars (MSPs, hereafter). This model has been quite 
successful in explaining and predicting several properties of binary MSPs
population with low mass, degenerate (Withe Dwarfs), companions
like their orbital period distributions and companion masses
(see {\it e.g.} Podsiadlowski, Rappaport \& Pfahl, 2002 for detailed
numerical orbital evolutions of short orbital period systems in which the
mass-transfer is driven by magnetic braking and/or gravitational radiation
(hereafter MB and GR, respectively),
and the classical work of Webbink, Rappaport and Savonije, 1983, for
semi--analytical evolution for long period systems driven
by stellar evolution of the companion off the main sequence branch;
see also Verbunt 1993 for a comprehensive review). However,
three observational facts were difficult to reconcyle with the
predictions of the Recycling Scenario: {\it i)} the lack of LMXBs showing
millisecond X--ray pulsations, {\it ii)} the fact that some
(up to 20\%) of MSPs are isolated, {\it iii)} the lack of NS spinning 
below $\sim 1.4 \ {\rm ms}$ (the shortest spin
period detected up to date: PSRJ1748--244ad, Hessels {\it et al.} 2006),
since the shortest spin periods attainable for NSs are well below 
one millisecond for most of the proposed equation of state of the
ultradense matter (see {\it e.g.} Lavagetto
{\it et al.} 2004).

In 1998, the discovery of coherent millisecond X--ray pulsations at
2.5 ms in \saxj, a transient LMXB observed, during an X-ray outburst,
by the {\it Rossi} X-ray Timing Explorer satellite (RXTE, hereafter)
solved {\it i)} by demonstrating that indeed some LMXBs
host a weakly magnetized, millisecond spinning NS (Wijnands \& van
der Klis 1998): the long sought progenitors
of MSPs were eventually found (the difficulty in detecting them was
the weak pulsed fracion, few percent, ascribed to the weakness of the
magnetic field strength, and the short orbital period of the binary,
about 2 h for \saxj, causing a further reduction of the power of the
coherent signal because of strong Doppler modulation).

Following this first discovery a total of 14
Accreting Millisecond X--ray Pulsar (AMP, hereafter) were discovered
to date, 
all in compact binaries ($P_{\rm ORB} < 1 \ {\rm d}$), 
all transient LMXB, few recurrent (see Table 1). 

In the following I will quickly reviev all the major progresses made,
in recent years in the topics outline above.
\section{Accretion in LMXBs: the fifteen golden years of AMPs}
A possible solution to {\it ii)} arrived after the 
discovery (Fruchter, Stinebring \& Taylor 1988)
of the original ``Black Widow'' B1957+20, a radio eclipsing 
MSP spinning at  $1.6 \ {\rm ms}$ in a close ($P_{\rm ORB}\sim 9.2 \ {\rm h}$) 
binary system with a very low mass companion 
($M_2 \sim 0.02 \ {\rm M_{\odot}}$).
It has been proposed that isolated MSPs have 
ablated their companions by means of 
low--freqency electromagnetic radiation, 
energetic particles and/or Gamma--rays produced by the fast 
rotating magnetic dipole (see {\it e.g.} Ruderman, Shaham
\& Tavani 1989). The huge NS rotational energy, stored
during the accretion phase, 
$E_{\rm ROT}/(Mc^2) =  0.011 \ I_{45}P_{-3}^{-2} m^{-1} \ {\rm erg}$,
is released according to the Larmor's formula:
$L_{\rm SD} = (2/3c^3) \mu^2 (2\pi/P^4) =
3.85 \times 10^{35} B_8^2 R_6^6 P_{-3}^{-4}\ {\rm erg/s}$, where
$M$ ($m=M/{\rm M_{\odot}}$),
$R$ ($R_6$ in units of $10^6$ cm), and
$I$ ($I_{45}$ in units of $10^{45} \ {\rm erg \ cm^2}$) are the NS 
mass, radius, and moment of inertia, respectively,
$P$ ($P_{-3}$ in units of $10^{-3}$ s) is the NS spin period,
$\mu = B R^3$ is the NS magnetic moment, 
$B$ ($B_8$ in units of $10^8$ Gauss) is the dipolar magnetic 
field strength at the NS magnetic equator, 
($\mu_{26} = \mu/(10^{26}\ {\rm Gauss \ cm^3}) = 1$ for $B_8 = R_6 = 1$),  
and $c$ is the speed of light. Thus, in case of no accretion,
the dipolar magnetic field strength at the NS equator is
$B_8 = 1.01 \ R_6^{-3}I_{45}^{1/2}(P_{-3} \dot{P}_{-20})^{1/2}$,
where $\dot{P}_{-20} = dP/dt/10^{-20}$.

To solve {\it iii)} the emission of GR from rapidly 
spinning NS has been proposed
(see {\it e.g.} Chakrabarty et al. 2003, Melatos \& Payne 2005 for models
of a non-zero mass quadrupole for an accreting pulsar), 
although, in most models,
the onset of rotational instabilities, that cause the emission of GR,
are predicted to occur at spin periods much shorter than the minimum observed
(see {\it e.g.} Burderi \& D'Amico, 1997).

Alternatively, to simultaneously explain {\it ii)} and {\it iii)},
our group proposed (Burderi et al. 2001) an evolutionary phase, 
that we dubbed ``Radio Ejection'' (RE, hereafter), that unavoidably 
triggers during accretion 
for systems with 
$P_{\rm ORB} \geq 
P_{\rm CRIT} \sim  1.05 \ {\rm h} \times 
L_{36}^{51/25} m^{1/10} \mu_{26}^{-24/5} 
P_{-3}^{48/5} \times  
\left[ 1 - 0.462 \left( {m_2} \over {m+m_2} 
\right)^{1/3} \right]^{-3/2} 
(m+m_2)^{-1/2}$,
where
$L_{36}$ is accretion luminosity in units of $10^{36} \ {\rm erg/s}$,
$m_2=M_2/{\rm M_{\odot}}$.
Indeed, since $P_{\rm CRIT} \propto P^{10} B^{-5}$, 
it quickly decreases below
$1 \ {\rm h}$ as $P$ decreases towards $1 \ {\rm ms}$ during the LMXB phase,
for any $B >  10^{8} {\rm \ Gauss}$. 
Which eventually causes the onset of RE for {\it any} system with 
$P_{\rm ORB} \geq 1 \ {\rm h}$. 
During this phase, the radiation pressure of the rotating magnetic dipole
($p_{\rm RAD}(R_{\rm RL}) = L_{\rm SD}/(4 \pi c R_{RL}^2)$, where $R_{\rm RL}$ 
is the distance 
of the Inner Lagrangian Point from the NS centre)
is sufficient to halt and expel the matter 
overflowing the inner Lagrangian point with a pressure
$p_{\rm RAM}(R_{\rm RL})= 0.5 \rho v^2 \sim 
\dot{M}\sqrt{GM/R}/(8\pi R_{\rm RL}^2)$. 
Similat to ablation, RE is more effective in ``evaporating'' 
the companion, since
it maximizes the fraction of $L_{\rm SD}$ intercepted by the companion, and 
minimize the specific binding energy of the matter that has to be ejected,
explaining {\it ii)}.    
Moreover, the $P^{10}$ dependence of $P_{\rm CRIT}$, acts as a barrier allowing
$P \ll 1.4 \ {\rm ms}$ only for extremely  compact systems
$P_{\rm ORB} \ll 1 \ {\rm h}$. Strong Doppler modulation
severely hampers the detection of a coherent signal in these cases,
explaining {\it iii)}. 

A spectacular confirmation of our model was the discovery of the eclipsing
pulsar PSRJ1740--5340 in the Globular Cluster NGC6397 ($P_3 = 3.65$,
$B_8 = 7.7$,$P_{\rm ORB} = 1.35 \ {\rm d}$, $m_2 \ge 0.19$, 
D'Amico et al. 2001) 
and the subsequent identification of its optical companion 
(Ferraro et al. 2001)
which showed evidence of ellipsoidal modulation and a size (determined
by the duration of eclipses) only consistent with a Roche Lobe 
filling companion.
Burderi, D'Antona, \& Burgay (2002) argued that PSRJ1740--5340 is 
in the RE phase
. 
Using the ATON1.2 code (D'Antona, Mazzitelli, \& Ritter 1989)  
we were able to effectively model the
binary evolution, considering the heating of the companion because 
of $L_{\rm SD}$,
successfully reproducing the peculiar Hertzsprung--Russel diagram location 
of the optical companion.

More recently, radio observations of Gamma--ray sources detected by
the Fermi satellite allowed to discover many (43 so far) MSPs,
with an extreme high fraction of eclipsing sources in compact binaries,
21 up to date. This has opened a unique opportunity to deeply study these 
systems, given that, in the 20 years from the discovery
of the original Black Widow B1957+20 (1990) up to the Fermi era (2009), 
only 3 eclipsing MSP were discovered (see Roberts 2012 for a review).
Eclipsing MSPs in compact systems ($P_{\rm ORB} \la 1 \ {\rm d}$)
were recently classified in ``Black Widows'' (14 out of 21, BW, hereafter) and 
``Redbacks'' (7 out of 21, after the Australian cousin of the 
poisonus American spider, 
RB, hereafeer), depending on the companion mass $\le 0.1 \ {\rm M_\odot}$ 
for BW, and $\ga 0.1 \ {\rm M_\odot}$. Companion masses are derived 
from optical
observations (Betron et al. 2013), indicating that, over a total of 17 BW and
7 RB, 21 have sizes $\ge 0.8$ of their Roche Lobes.

We argue that: a) PSRJ1740--5340 can be considered the
first discovered (prototype) RB, b) BW and RB are almost all in RE phase.
Otherwise, if the companion underfills its Roche Lobe, and no mass--transfer
occurs, it would be difficult to imagine a mechanism to replenish the plasma 
responsible for the eclipses, as discussed in Breton et al. (2013). 
Our evolutionary scenario predicts this, 
since we showed that the onset of RE is the natural outcome for 
any system with $P_{\rm ORB} \geq 1 \ {\rm h}$.

The discovery of AMPs offered the unique opportunity to
verify, during the accretion phase, 
the evolutionary path outlined above. 
This has been made possible thanks to the compelling opportunity
of applying the accurate machinery of timing analysis to millisecond
``clocks'' moving in close orbits. 
In particular the astonishing precision of these tecniques allowed to 
highlight, at least in some cases, the istantaneous effect of the 
accretion of angular momentum onto the NS, through an accurate measure 
of the NS spin--up rate.
This is a strong verification of several underlying hypothesis usually
postulated in the Recycling Scenario.

Given the NS spin--up rate, adopting a reasonable prescription for the torque
exerted by the accretion flow ({\it e.g.} that the matter accretes its
specific angular momentum at the point where is captured by the NS
magnetic field), it is possible to verify the presence of matter orbiting
in a Keplerian accretion disc truncated at the magnetospheric radius
$R_{\rm m}$, comparable to the Alfv\'en radius. This is the radius
where the energy density of the NS magnetic dipole is equal to the
energy density of the accretion flow, assumed in free fall:
$R_{\rm A}= (2GM)^{-1/7} \dot{M}^{-2/7} \mu^{4/7} = 9.90 \times 10^5 m^{-1/7}
\dot{m}_{-8}^{-2/7} (B_8 R_6^3)^{4/7}\; {\rm cm}$, where $G$ is the gravitational
constant, $\dot{M}$ ($\dot{m}_{-8}=\dot{M}/(10^{-8}{\rm M_{\odot}}/{\rm yr})$) 
is the accretion rate.

Moreover, since the dynamical effect of accretion is measured directly
through the spin--up rate, this independent estimate of $\dot{M}$ can be
compared with the X--ray (almost bolometric) luminosity 
($L = \eta GM\dot{M}/R$)
to infer, given the distance of the source, the efficiency of accretion.
The theoretical prediction $\eta \sim 1$ has been experimentally confirmed
(see {\it e.g.} Burderi {\it et al.} 2007, Papitto {\it et al.} 2008).

A clear determination of the spin--up rate during the x--ray outburst 
is {\it per se} very complex since the expected spin period derivatives 
are tiny ($\dot{P} \sim 10^{-18}$) (for $\dot{m}_{-8} \sim 0.1$ corresponding 
to typical X--ray outburst luminosities of $10^{37} \ {\rm erg/s}$) 
acting for just few tens of days (typical outburst duration). 
This implies variations for the phases of the pulse profile 
$\Delta \phi / \phi \la 1$. Unfortunately in some cases,
phases are affected by strong timing noise with 
$\Delta \phi_{\rm NOISE} / \phi \la 1$.  
By expanding the almost sinusoidal pulse profile in a Fourier Series
(no more than 4 harmonic components are typically needed, and only the 
first two are significant in most cases) we performed timing analysis
on each component and discovered that, in most cases, the second harmonic
component is noise--free and shows a nearly parabolic trend in time, which 
is expected for a constant spin--up torque (see {\it e.g.} 
Burderi {\it et al.}
2006). We therefore argued that the phases of second harmonic component are 
a more stable tracer of the NS spin evolution.

To take into account the variations of the accretion rate along the 
outburst and the threading of the accretion disc by the NS magnetic 
field, we adopted the torque prescription of Rappaport, Fregeau, 
\& Spruit (2004): 
$\tau_{\rm NS} = \dot{M}\sqrt{GMR_{\rm CO}} - \mu^2/(9R_{\rm CO}^3)$,
where $R_{\rm CO} = 1.5 \times 10^6 \ m^{1/3}P_{-3}^{2/3} \ {\rm cm}$ 
is the corotation radius (at which the speed of NS magnetic field lines 
is equal to the local Keplerian speed in the disc). We were able to  
perform timing analysis on 7 out of 14 sources in which
timing noise was either absent ({\it e.g.} \igra, Burderi et al. 2007) or
strongly suppressed in the second harmonic component ({\it e.g.} \xtea,  
Riggio {\it et al.} 2008). In 5 cases spin--up consistent with the 
expectations of the Recycling Scenario was detected, while in 2 cases the
detected spin--down suggest the presence of a strong NS magnetic moment
({\it cf.} the torque formula, above). 
The presence of the timing noise originally raised doubts on the reliability
of the spin period derivatives determined with timing analysis 
(Hartman et al. 2008). Heterodox phenomenological models in which pulse phases
correlate with X--ray luminosities were also proposed (Patruno, Wijnands, \&
van der Klis 2009). However, at least for the cases in which weak timing noise
was present, there is now general consensus that phases are a good tracer 
of the NS spin (see {\it e.g.} Patruno 2010).

In the few cases in which the AMPs are recurrent, timing analysis of all the
X--ray outbursts allows to compute the secular evolution of orbital parameters
and, in particular, of the orbital period. This can be compared with the
theoretical predictions derived from the assumption of an orbital evolution
driven by angular momentum losses determined by GR and/or MB,
as expected in these systems. This has been done for \saxj, 
in which we detected orbital expansion at an almost constant rate more than 
10 times what is expected by conservative mass transfer from a fully 
convective 
and/or degenerate secondary driven by the emission of GR 
(Di Salvo et al. 2008, Hartman et al. 2008). We performed orbital evolution
calculations demonstrating that highly non-conservative (about 99\%) 
mass--transfer could explain the observed orbital expansion. 
In our model mass transfer proceed at almost constant rate, 99\% of 
this matter is 
expelled during quiescence and 1\% accreted during recurrent outbursts. 
Severe angular
momentum losses, caused by the ejected matter, speed--up orbital evolution
increasing mass--transfer rates with respect to those expected by the action 
of GR (Di Salvo et al. 2008, Burderi et al. 2009).
Therefore we argued that \saxj\ is alternating between RE and accretion 
episodes,
which is expected if evolution has driven the system in a phase in which
$p_{\rm RAD}(R_{\rm RL}) \sim p_{\rm RAM}(R_{\rm RL})$.
Our evolutionary scenario, outlined above, predicts this behaviour 
as an equilibrium endpoint of the mass--transfer phase in LMXBs harbouring
a magnetized fast spinning NS. Therefore we believe that most of the AMP 
alternate
between accretion and RE phases.

Based on a small increase of the orbital period expansion during the 
last (2011) 
outburst, Patruno et al. (2012) concluded that the most plausible explanation 
for orbital period accelerated expansion was given by a companion 
spin--orbit coupling 
(see Applegate \& Shaham 1994 for a description of the model).
Alternatively we argue that the observed acceleration of the orbital expansion 
with respect to the constant expansion rate is a necessary consequence of a 
doubling of the mass-transfer, demonstrated by the doubling of the X--ray flux 
observed in the 2011 outburst (Burderi {\it et al.},
{\it in preparation}).
 
Given the huge amount of power emitted by the fast spinning
magnetic dipole (up to 100 $L_\odot$, see above), it is conceivable
that during X--ray quiescence, when this luminosity is not overwhelmed
by the accretion power, irradiation effects on the face of the companion
exposed to the NS dipole radiation, could manifest in a detectable
modulation at the orbital period. The attractive idea is to use the companion
as a bolometer to chatch the albeit elusive magnetodipole radiation in all the
energetic channels in which it is emitted: low frequency ($\nu = 1/P)$
electromagnetic waves, $\gamma$ rays, $e^{+/-}$ pais, etc. From the reasonable
assumption that the companion fills its Roche Lobe, the fraction of power
intercepted can be accurately determined, which allows to estimate the power
emitted by the rotating magnetic dipole quite accurately, and therefore 
to infer
an independent estimate of the NS magnetic field strength. 
Burderi {\it et al.} 
(2003) applied this idea to interpret the optical luminosity of the companion 
of \saxj \ during quiescence, which was overluminous for $m_2 \le 0.18$ 
(Chakrabarty \& Morgan 1998). We found that optical data implied 
$1 \le B_8 \le 5$. 
This method was subsequently applied with success
by several authors (see {\it e.g.} Campana {\it et.al.}2004, 
D'Avanzo {\it et.al.} 2009).

From what is discussed above, it is clear that a crucial role in the whole 
Recycling Scenario is played by the magnetic moment of the NS. This is the 
means by which a fraction of the NS huge mechanical energy (slowly 
accumulated during the previous, long lasting, accretion phase) is released 
in the system, altering dramatically its entire evolution.
Therefore we finally discuss the estimates of the magnetic field of AMP, 
through four independent methods, namely: a) X--ray residual 
luminosity in quiescence
(Di Salvo \& Burderi 2003), b) optical reprocessing of 
rotating magneto--dipole 
radiation, discussed above, c) fitting pulse phase delays 
with the torque formula
discussed above, d) for AMPs showing recurrent outburts, comparing subsequent
spin period estimates to infer the secular spin--down induced by $L_{\rm SD}$. 
%
%
\begin{acknowledgements}
The authors want to warmly thank Franca D'Antona for the long lasting 
fruitful collaboration which has lead many compelling discoveries in 
this exciting field. 
Her profound knowledge of the evolution of stars in binary
systems has revealed to be an unvaluable ingredient for understanding 
these fashinating systems.
We like to remember several thrilling discussions during which 
unconventional possibilities were thorougly discussed because
{\it ``when you have eliminated the impossible, whatever remains, however improbable, must be the truth} 
(Sherlock Holmes, by Sir Arthur Conan Doyle).
We also thank the organizers for the nice and interesting conference 
and the warm hospitality in Rome Astronomical Observatory.
\end{acknowledgements}
\begin{table*}
\caption{AMP parameters. Notes: 
$^{(1)}$ adopting $M_1 = 1.4 \ {\rm M}_{\odot}$,
and $\sin i = 90^{\circ}$;
$^{(2)}$ during X--ray outbursts;
$^{(3)}$ when derived form secular spin down
$B_8 = 1.01 \ R_6^{-3}I_{45}^{1/2}(P_{-3} 
\dot{P}_{-20})^{1/2}$ with $R_6 = I_{45} = 1$ is adopted;
$^{(4)}$ gal05:  Galloway    {\it et al.} 2005;  
          cas08:  Casella     {\it et al.} 2008;
          alt11:  Altamirano  {\it et al.} 2011;
          alt08:  Altamirano  {\it et al.} 2008;  
          mar02:  Markwardt   {\it et al.} 2002;
          wij98:  Wijnands \& van der Klis 1998;
          pap11b: Papitto     {\it et al.} 2011b;
          kar06:  Kaaret      {\it et al.} 2006;  
          mar03:  Markwardt \& Swank       2003;
          mar09:  Markwardt   {\it et al.} 2009;
          alt10:  Altamirano  {\it et al.} 2010;
          rig08:  Riggio      {\it et al.} 2008; 
          gal02:  Galloway    {\it et al.} 2002;
          kri07:  Krimm       {\it et al.} 2007;
$^{(5)}$ Papitto {\it et al.} 2011a, based on the 10 days duration 
of the 2004 outburst 
($|\dot{P}_{-18}| \le 2.3$ at 3 $\sigma$ level, Galloway {\it et al.} 
2005, based on the first 3 days of data for a total burst duration of 10 days;
$\dot{P}_{-18} = - 2.43 \pm 0.17$, Falanga {\it et al.} 
2005, based on the last 7 days of data; 
$\dot{P}_{-18} = - 2.46 \pm 0.32$, Burderi {\it et al.} 
2007, based on the last 7 days of data; 
$\dot{P}_{-18} = - 1.62 \pm 0.09$, Patruno 2010, based on the 10 days 
duration of 
the 2004 outburst);
$^{(6)}$ Papitto {\it et al.} 2011a, method (d) \ ($B_8 = 1.05 \pm 0.14$, 
Patruno 2010, method (d);$B_8 = 1.42 \pm 0.21$, 
Hartman {\it et al.} 2011, method (d));
$^{(7)}$ Papitto {\it et al.} 2008;
$^{(8)}$ Riggio {\it et al.} 2011b, method (d);
$^{(9)}$ Chakrabarty \& Morgan 1998;
$^{(10)}$ Burderi {\it et al.} 2006;
$^{(11)}$ Patruno {\it et al.} 2012, method (d)\ 
($B_8 = 0.93 \pm 0.42$, Hartman {\it et al.} 2009, method (d); 
$1 \le B_8 \le 5$, Di salvo \& Burderi 2003, method (a); 
$1 \le B_8 \le 5$, Burderi {\it et al.} 2003, method (b); 
$B_8 \ge 0.6$, Campana {\it et al.} 2004, method (b); 
$B_8 = 3.5 \pm 0.5$, Burderi {\it et al.} 2006, method (c) 
from timing of the first harmonic. 
Encouragingly, for \saxj \ the four methods are
consistent within each other at less than $3 \sigma$ level);
$^{(12)}$ Papitto {\it et al.} 2007 from timing of the fundamental 
($P_{18}= 0.87 \pm 0.09$ from timing of the first harmonic);
$^{(13)}$ Papitto {\it et al.} 2007, method (c);
$^{(14)}$ Riggio {\it et al.} 2011a, from timing of the first harmonic;
$^{(15)}$ Galloway {\it et al.} 2002, method (a).
}
\label{amptable}
\begin{center}
\begin{tabular}{lcrlccl}
\hline
\hline
\\
${\rm Source \; name}$    & $P_{-3}$ & $P_{\rm ORB} \ {\rm (h)}$ & 
$m_{2\;{\rm min}}^{(1)}$ & $\dot{P}_{-18}^{(2)}$  & $B_8^{(3)}$ &         
${\rm reference^{(4)}}$ \\ 
\\
IGR J00291+5934 & $1.7$ & $2.5$ & $0.039$ &$-1.47 \pm 0.09^{(5)}$ & $1.43 \pm 0.21^{(6)}$ & 
gal05 \\
Aql X--1 (*)    & $1.8$ & $19$ & -- & -- & -- & 
cas07 \\
SWIFT J1749.4-2807 & $1.9$ & $8.8$ & $0.59$ & -- & -- & 
alt11 \\
SAX J1748.9-2021 & $2.3$ & $8.8$ & $0.1$ & -- & -- & 
alt08 \\
XTE J1751-305 & $2.3$ & $0.7$ & $0.014$ & $-1.96 \pm 0.53^{(7)}$ & $2.61 \pm 0.28^{(8)}$ & 
mar02 \\
SAX J1808.4-3658 & $2.5$ & $2.0^{(9)}$ & $0.043^{(9)}$ & $-2.75 \pm 0.51^{(10)}$ & 
$1.62 \pm 0.10^{(11)}$ & 
wij98 \\
IGRJ17498-2921 & $2.5$ & $3.8$ & $0.17$ & $+0.39 \pm 0.12$ & -- & 
pap11b \\
HETE J1900.1-2455& $2.7$ & $1.4$ & $0.016$ & -- & -- & 
kar06 \\
XTE J1814-338 & $3.2$ & $4.3$ & $0.15$ & $+0.69 \pm 0.07^{(12)}$ & $\sim 8^{(13)}$ & 
mar03 \\
IGR J17511-3057 & $4.1$ & $3.5$ & $0.13$ & $-2.44 \pm 0.27^{(14)}$ & -- & 
mar09 \\
NGC 6440 X-2 & $4.8$ & $0.95$ & $0.0067$ & -- & -- & 
alt10 \\
XTE J1807-294 & $5.2$ & $0.67$ & $0.0067$ & $-0.68 \pm 0.19$ & -- & 
rig08 \\
XTE J0929-314 & $5.4$ & $0.73$ & $0.008$ & $+2.68 \pm 0.12$ & $\sim 10^{(15)}$ & 
gal02 \\
SWIFT J1756.9-2508 & $5.5$ & $0.9$ & $0.007$ & $|\dot{P}_{-18}|< 10$ & -- & 
kri07 \\
\\
\hline
\hline
\end{tabular}
\end{center}
\end{table*}
\bibliographystyle{aa}

\end{document}